\documentstyle[12pt,aasms4]{article}
\begin{document}

\title{HCG 16 Revisited: Clues About Galaxy Evolution in Groups}

\author{Reinaldo R. de
Carvalho\altaffilmark{1}, Roger Coziol\altaffilmark{1,2} } 

\altaffiltext{1}{Observat\'orio Nacional, Rua Gal. Jos\'e Cristino,
77 -- 20921-400, Rio de Janeiro, RJ., Brasil}
\altaffiltext{2}{PRONEX/FINEP -- P. 246 -- 41.96.0908.00}

\begin{abstract}

We present new spectroscopic observations of 5 galaxies, members of the 
unusually active compact group HCG 16, observed using the 
Palomar 5m telescope. The high signal to noise ratios (S/N $\sim 70$)
of the spectra allow us to study the variation of the emission line characteristics
and the stellar populations in the nucleus and the circumnuclear regions
of the galaxies. 
The emission line characteristics of these galaxies are complex, 
varying between Seyfert 2 and LINERs or between 
LINERs and starbursts. All of the galaxies show traces of intermediate
age stellar populations, supporting our previous result 
that post-starburst galaxies are common in compact groups. 
The galaxies HCG16--4 and HCG16--5 show double nuclei and therefore could be
two cases of recent merger.

Our observations support a scenario where HCG 16 was formed by the successive
merger of metal poor, low mass galaxies. The galaxies HCG16--1 and HCG16--2, 
which are more
evolved, form the old core of the group. Galaxies HCG16--4 and HCG16--5 are two more 
recent additions still in a merging phase. Galaxy HCG16--5 is a starburst galaxy
which is just beginning to fall into the core. If HCG 16 is representative
of compact groups in their early stage, the whole set of observations implies that 
the formation of compact groups is the result of
hierarchical galaxy formation. HCG 16 could be one example of this process
operating in the local universe.  

\end{abstract}


\section{Introduction}

To study the dynamical structure of compact groups 
of galaxies, de Carvalho et al. (1997) obtained new spectroscopic data 
on 17 of Hickson's compact groups (HCGs), extending the observations to 
galaxies which are in the immediate vicinity of the original 
group members (within 0.35 Mpc, H$_{\circ}$= 75 km/s/Mpc, from the nominal center,
in average, Ribeiro et al. 1998). The analysis based on this
survey (Ribeiro et al. 1998; Zepf et al. 1997) helped 
to resolve some of the ambiguities presented by the HCGs. 
In particular, it revealed that compact groups may be different dynamical stages of evolution of
larger structures, where replenishment by galaxies from the halo is always 
operating. Several other papers have addressed this particular scenario from either
the observational or theoretical point of view (e.g. Barton et al. 1996; 
Ebeling, Voger, \& Boringer 1994; Rood \& Struble 1994; Diaferio, Geller, \&
Ramella 1994, 1995; Governato, Tozzi, \& Cavaliere 1996). 

Consistent with the dynamical analysis, the classification of the 
activity types and the study of the stellar populations of the galaxies in these groups 
suggest that their evolution followed similar paths and that they were 
largely influenced by their environment (Ribeiro et al. 1998; Mendes de Oliveira et al. 1998).
Most of the groups have a core (basically corresponding to the Hickson definition of the group)
 and halo structure (see Ribeiro et al. 1998 for a definition of the halo population).
The core is dominated by AGNs, dwarf AGNs and galaxies whose spectra do not show any emission, whereas  
starbursts populate the halo. 
The AGNs are located in the most early--type, luminous galaxies and are
preferentially concentrated towards the central parts of the groups.
The starbursts in the halo, on the other hand, appear to be located preferentially
in late--type spiral galaxies (Coziol et al. 1998a, 1998b).
This last result for the core of the groups was 
recently confirmed by Coziol et al. (1998c) from a study of 
a new sample of 58 compact groups in the southern hemisphere (Iovino \& Tassi 1998). In this
study, we also show that no Seyfert 1s have been found in out sample of compact groups.

In terms of star formation and populations, the galaxies in the core of
the groups (the ``non--starburst'' galaxies) seem more evolved than those 
in the outer regions: the galaxies are more massive
and more metal rich than the starbursts and they show little or no star formation. 
Most of these galaxies have, however, stellar metallicities which are unusually
high compared to those of normal galaxies with similar morphologies (Coziol et al. 1998b).  
They also show unusually narrow equivalent widths of metal absorption lines
and relatively strong Balmer absorption lines, which are consistent with the presence
of a small (less than 30\%) population of intermediate age stars (Rose 1985). These 
observations suggest 
that most of the non--starburst galaxies in the groups are in a relatively 
evolved ``post-starburst'' phase (Coziol et al. 1998b). 

HCG 16 is a group composed of 7 galaxies with a mean velocity V$ = 3959 \pm 66$ km s$^{-1}$ and 
a dispersion $\sigma = 86 \pm 55$ km s$^{-1}$ (Ribeiro et al. 1998). Although we are keeping
Hickson's nomenclature for this group, it is important
to note that we are not following specifically Hickson's definition of a group, since
this is not a crucial point for our analysis. Besides, there is evidence that HCG 16 is
part of a larger and sparser structure (Garcia 1993). Specific studies have been done on
HCG 16, covering a broad domain of the electromagnetic spectrum, allowing a thorough exam
of its physical properties. Radio and infrared (Menon 1995; Allam et al. 1996); CO observations
estimating the mass of molecular gas in some of the HCG16's members (Boselli et al. 1996);
rotation curves exhibiting abnormal shapes (Rubin, Hunter, \& Ford 1991). Hunsberger et al. (1996)
detected some dwarf galaxy candidates for HCG16-a, which is interpreted as a sign of strong 
interaction.
From the spectral characteristics, Ribeiro et al. (1996) identified one Seyfert 2 galaxy, two LINERs 
and three starburst galaxies. Considering the significant amount of information gathered for
HCG 16, this group represents a unique opportunity to
obtain new clues on the process of formation of the compact groups. Here in this paper we focus on
study of the activity of five galaxies belonging to the group: four galaxies originally
defined as the Hickson group number 16 and the fifth one added from Ribeiro et al. (1998).
These authors re-defined this structure with seven galaxies (including the original four from
Hickson), but we gathered high quality data for only five of them.

\section{Observations and data reduction}

Spectroscopic observations were performed at the Palomar 200-inch telescope using
the Double Spectrograph on UT 1996 October 16. Typical exposure times were 600 to 900
seconds depending on the magnitude of the galaxy. Two gratings were used: one for the
red side (316 l/mm, resolution of 4.6 \AA), and one for the blue side (300 l/mm, resolution of 4.9 \AA). 
The wavelength coverage was 3800\AA\ to 5500\AA\ in the blue and 6000\AA\ to 8500\AA\ in the red. 
For calibration, He--Ne arc lines were observed before and after each exposure throughout the night.
During the night, the seeing varied around 1.5 arcsecs.  
It is important to stress that in this paper we present only a qualitative discussion of the 
relative rates of star formaton since the data were taken under non-photometric conditions 
hampering a proper flux calibration.

The reduction of the spectra was done in IRAF using standard methods.
An overscan was subtracted along the dispersion axis, which took care of the bias correction. 
All the spectra were trimmed and divided by a normalized flat field.
Wavelength calibration, done through a polynomial fit to the He--Ne arc lines, gave 
residuals of $\sim$0.1\AA.

The relatively high signal to noise ratios of the spectra (S/N $\sim70$ on average), 
allow us to study the variation of the emission line characteristics and
stellar populations as a function of their position in the galaxies. 
To do so, the reduction to one dimension was done in the case of the red spectra  
using up to 7 apertures of $\sim3$ arc seconds in width. Due to the lower S/N level obtained, 
only 3 apertures were used in the blue part of the
spectrum. To compare the line ratios and absorption features
in the red with those measured in the blue, the reduction
was also redone in the red using only 3 apertures. 

In the case of the spectra reduced with 3 apertures,  
the spectrum of the galaxy NGC 6702 was used as a template to correct
for contamination by the stellar populations (Ho 1996, Coziol et al. 1998a). 
Before subtraction, the spectrum of the template was 
normalized to fit the level of the continuum in the galaxies and in one case,
HCG 16--5, the Balmer absorption lines were artificially enlarged to fit the 
broad absorption lines observed in this galaxy. 

\section{Results}

\subsection{Distribution of the light and ionized gas in the spectra} 

Table 1 gives the basic characteristics of the 5 galaxies studied in this paper.
The numbers in column 1 follow the nomenclature used in Ribeiro et al. (1996). 
The radial velocities in column 2 and the absolute magnitudes in
column 3 were taken from Coziol et al. (1998b). 
The morphological types listed 
in column 4 were taken from Mendes de Oliveira \& Hickson (1994). The
different types of activity in column 5 correspond to our new classification
as presented in Section 4 and Figure~3. The complexity of the AGNs is obvious from the
multiple characteristics of their spectra. 
The next 3 columns correspond to the extension of the 
projected light on the spectra, as deduced from the red part of the spectrum.
The total galaxy is measured from the extension until the signal reaches the sky level.
The ionized region corresponds to the projected length where emission can be seen.
The nucleus corresponds to the extension of light at half maximum intensity (FWHM).
With the exception of HCG16--1, all the galaxies have a nucleus which is well resolved.    
The last column gives for each galaxy the equivalent of 1 arc second in parsecs.

Figure~1 shows, on the left, the extension of the ionized gas, as traced by H$\alpha$ and 
the two [\ion{N}{2}] lines, and, on the right, the light profile along
the slit. In the galaxies HCG16--1, HCG16--2 and HCG16--3, 90\% of the light is concentrated 
in a window $\sim 9$ arcsecs wide, which corresponds to $\sim 2$ kpc at the distance of the galaxies.
The remaining 10\% of the light extends over a region not exceeding 8 kpc. 
These galaxies look compact compared to normal spiral galaxies.

In galaxies HCG16--4 and HCG16--5 the light is slightly more extended ($\sim 3$ and 6 kpc, respectively), 
but this is because these two galaxies
probably have a double nucleus. The second nucleus in HCG16--4 corresponds
to the second peak 5 arcsecs  west of the primary nucleus, while in HCG--5
the second nucleus corresponds to the small peak 7 arcsecs east of the primary nucleus.
It is very unlikely that these structures could be produced by dust,
because we are using the red part of the spectra where 
extinction effects are minimized. 
In the next section, we will show also that the second nucleus in HCG16--5 presents a 
slightly different spectral characteristic compared to the primary nucleus, which is inconsistent with
the idea that this is the same galaxy.   
HCG16--4 and HCG16--5 are probably the product of recent mergers of galaxies. Other
studies present strong evidence of central double nuclei (Amram et al. 1992; Hibbard 1995). 

In all the galaxies, the ionized gas is more intense and mostly concentrated in the nucleus.
\ion{H}{2} regions outside the nucleus are clearly visible only in HCG16--1 and HCG16--3.  
It looks like the activity (star formation or AGN) is always concentrated in the center
of the galaxies. In HCG16--5, the second nucleus seems less
active (we see less ionized gas) than the primary nucleus, while in HCG16--4, the two nuclei
appear equally active. 

\subsection{Variation of the activity type with the radius} 

In Ribeiro et al. (1996) we already determined the activity types of 
these galaxies. Having in hand spectra with high
S/N we now repeat our analysis of the activity for the five most
luminous galaxies, but this time separating each spectrum in various
apertures covering different regions in order to see how activity varies 
with the radius.  

In Figure~2, we present the results of our classification of the activity type 
using the standard diagnostic diagram (Baldwin, Phillips \& Terlevich 1981;
Veilleux \& Osterbrock 1987). The line ratios correspond to the values obtained after
subtraction of the template galaxy NGC 6702. Because of the relatively lower S/N of the blue
as compared to the red part of the spectra, we limit our study to only three apertures. 
In Figure~2, the first apertures, identified by filled
symbols, cover the nucleus. The two other apertures cover regions to the east and 
to the west of the nucleus.
The width of these apertures can be found in column~3 of Table~3. 
Note that these apertures are covering mostly the central part of the galaxies.   

Our new classification is similar to the one given in Ribeiro et al. (1996).
In particular, the galaxies keep their original classification as an AGN or a starburst.
We note, however, some interesting variations. The most obvious of these 
variations concerns HCG16-1, which was classified as a luminous Seyfert 2 and now 
appears as a LINER nucleus with outer regions in a starburst phase.  
Another difference with our previous classification is related
to the discovery of the second nucleus in HCG16-5, although we do not find any evidence
of difference in excitation state of both nuclei, considering the large error bars (See Figure 2).  
We see very little variation in the other three galaxies.   
The level of excitation for HCG16-3 is higher suggesting that the gas in this
galaxy is slightly less metal rich than in HCG16-4 (McCall, Rybsky, \& Shields 1985; Evans \&
Dopita 1985). 

To study the variation of the activity in greater detail, we have 
divided the spectra in the red into 7 equal apertures of $\sim3$ arc seconds in width. 
In Table~2, the different apertures are identified by a number which increases from east to
west. The apertures centered on the nuclei are identified
with a small {\it n} and the circumnuclear regions with a small {\it ci}.
In column 3, the corresponding radius in parsecs is also given. 
The parameters that were measured are: the FWHM of the H$\alpha$ emission line (column 4) and the
ratio [\ion{N}{2}]$\lambda6548$/H$\alpha$ (column 5), which allow 
to distinguish between starbursts and AGNs (Baldwin, Phillips, \& Terlevich 1981; Veilleux
\& Osterbrock 1987; Ho, Fillipenko, \& Sargent 1993; V\'eron, Gon\c calvez, \& V\'eron-Cetty
1997); the equivalent width of H$\alpha$ (column 6), 
which in a starburst is a good indicator of the strength of star formation (Kennicutt 1983; 
Kennicutt \& Kent 1983; Copetti, Pastoriza, \& Dottori 1986; Salzer, MacAlpine, \& Boroson 1989;
Kennicutt, Keel, \& Blaha 1989; Coziol 1996); and the
ratio [\ion{S}{2}]$\lambda6716+\lambda6731$/H$\alpha$ (column 5), which we use as a tracer of the level
of excitation (Ho, Fillipenko, \& Sargent 1993; Kennicutt, Keel, \& Blaha 1989; Lehnert \&
Heckman 1996; Coziol et al. 1999). All the lines were measured using the standard routines in SPLOT, 
fitting the continuum by eye. 
A gaussian profile was usually assumed, though in some cases, a lorentzian was
used. The uncertainties were determined by comparing values obtained by measuring the
same lines in two different spectra of the same object. 

In Figure~3, we present the diagrams of the ratio [\ion{N}{2}]$\lambda6548$/H$\alpha$ as a function
of the EW of H$\alpha$. The corresponding regions are identified by their number in Table~2.
In these diagrams, AGNs usually have 
a higher [\ion{N}{2}]/H$\alpha$ ratio than starbursts, but smaller EW (Coziol et al 1998b).
We now examine each galaxy separately.

In HCG16-1, the star formation in the outer regions, as noted in Figure~2, appears quite clearly. 
As compared to HCG16-4, which is the strongest starburst we have in the group, the
relatively lower EW of these \ion{H}{2} regions suggests milder
star formation. The EW of H$\alpha$ is a measure of current to past star formation, the
relatively lower EW suggests, therefore, an older phase of star formation (Kennicutt, Keel, 
\& Blaha 1989; Salzer, MacAlpine, \& Boroson 1989; Coziol 1996).
The star formation is constant on the east side
of the galaxy (apertures 1 and 2) but decreases to the west (from apertures 6 to 7).
The nucleus and circumnuclear regions do not show any variation, the condition
of excitation of the gas staying constant out to a radius of $\sim 1.2$ kpc.

In HCG16-2, no star formation is observed. We see a slight variation in the 
circumnuclear regions, within a 1 kpc radius of the nucleus, and
a more significant variation in the outer regions. If we assume that
the source of the gas excitation is limited to the nucleus, the variation of 
the [\ion{N}{2}]/H$\alpha$ and EW in the outer regions can be explained by 
a simultaneous decrease of the gas excitation (H$\alpha$ flux goes down) and 
a change towards older stellar populations (EW H$\alpha$ decreases). 
This suggests that HCG16-2 is an AGN located in
a galaxy dominated by intermediate and older age stellar populations. In
starburst galaxies, the ratio [\ion{N}{2}]/H$\alpha$ is also sensitive to
the abundance of nitrogen (Evans \& Dopita 1985; Coziol et al. 1999). The increase
of [\ion{N}{2}]/H$\alpha$ in the outer regions, therefore,  could also suggests
an increase of the abundance of nitrogen (Stauffer 1982; Storchi-Bergmann 1991;
Storchi-Bergmann \& Wilson 1996; Ohyama, Taniguchi \& Terlevich 1997; Coziol et al. 1999). 
It may suggest a previous burst of star formation in the recent past of this AGN (Glass \& Moordwood
1985; Smith et al. 1998).

HCG16-3 is a starburst galaxy at the periphery of the four other
luminous members of HCG 16 and the only one in our sample which is not original member
of the Hickson group. Comparison with HCG16-4 indicates that the star formation
is at a lower level. Again, no variation is observed within $\sim 1.2$ kpc of the nucleus
while the [\ion{N}{2}]/H$\alpha$ ratio increases and EW decreases in the outer regions.
However, the variation of these two parameters is less severe than in the case of 
HCG16-2. Because HCG16-3 is classified as a starburst, we assume that the source of gas ionization 
is not limited only to the nucleus but follows the star formation.
The variation observed  would then mean that the star formation
in the outer regions (aperture 2 and 6) is at a more advanced stage of evolution than 
in the nucleus.  

The same behavior as in HCG16-3 is observed in HCG16-4. The star formation
in this galaxy, however, is at a more intense level.
This is probably because HCG16-4 is in a merger phase since this galaxy has a double nucleus. 
Contrary to HCG16-3, we see
also some spectral variations in the nucleus, consistent with
a double nucleus: apertures 3 and 2 correspond to the second nucleus
while apertures 4 and 5 correspond to the primary nucleus. Again the outer regions
seem to be in a more advanced stage of evolution than in the nucleus.  

The variations observed in HCG16-5 are much more complex than in the other galaxies. 
The presence of a second nucleus makes the interpretation even more difficult. 
In Figure 3, the second nucleus corresponds to apertures 6 and 7. 
It can be seen that the two nuclei have the same behavior. The variation of the parameters out
of the nuclei is similar to what we observed in the two starbursts HCG16-3 and HCG16-4, but the range 
of variation is more similar to that observed in HCG16-2. 
Although HCG16-5 was classified as a LINER, its nature seems ambiguous, showing
a mixture of starburst and AGN characteristics.
It is important to note the difference with respect to HCG16-1, which is 
a central AGN encircled by star
forming regions. In HCG16-5, on the other hand, the AGN in the nucleus seems 
to be mixed with intense star formation
(Maoz et al. 1998; Larking et al. 1998).
Out of the nucleus, there is no star formation and the AGNs may be responsible for  
ionizing the gas (Haniff, Ward, \& Wilson 1991; Falcke, Wilson, \& Simpson 1998; Contini 1997).  
   
\subsection{ Variation of the excitation with the radius}

Comparing the ratio [\ion{N}{2}]$\lambda6548$/H$\alpha$ with
the ratio [\ion{S}{2}]$\lambda6716+\lambda6731$/H$\alpha$ it is possible to distinguish 
between the different source of excitation of the gas (Kennicutt, Keel, \& Blaha 1989,
Ho, Fillipenko, \& Sargent 1993, Lehnert \& Heckman 1996). Shocks from surpernovae remnants 
in a starburst, for example, produce a [\ion{S}{2}]/H$\alpha$ ratio higher than 
0.6, much higher than the mean value of $\sim 0.25$ usually observed in normal 
\ion{H}{2} regions or in starbursts (Greenawalt \& Walterbos 1997; Coziol et al. 1997). 
In AGNs, however, the effect of shocks are more difficult to distinguish because 
both of these lines are highly excited (Baldwin, Phillips \& Terlevich 1981;
Veilleux \& Osterbrock 1987; Ho, Fillipenko, \& Sargent 1993; 
Villar-Mart\' {\i}n, Tadhunter, \& Clark 1997; Coziol et al. 1999). 
We will assume here
that a typical AGN has [\ion{N}{2}]/H$\alpha > 1$ and [\ion{S}{2}]/H$\alpha > 0.6$.

In Figure~4, we now examine the behavior of these ratios as a function
of the radius for each of the galaxies. In HCG16-1, although we
now classify the nucleus as a LINER, the  
values of the two ratios are still consistent with those of a typical AGN. 
The [\ion{N}{2}]/H$\alpha$ ratio for the outer starbursts are at the lower limit
of the value for AGNs, but the [\ion{S}{2}]/H$\alpha$ ratio is normal for
gas ionized by hot stars. On the other hand, the outer region 
corresponding to aperture 7 has a very unusually high ratio, which suggests
that this region could be the location of shocks (Ho, Fillipenko, \& Sargent 1993;
Lehnert \& Heckman 1996; Contini 1997). 

In HCG16-2, both ratios are high, consistent with its AGN nature. 
We note also that in the outer regions the [\ion{S}{2}]/H$\alpha$ ratio decreases 
or stays almost constant while the [\ion{N}{2}]/H$\alpha$ ratio increases. 
This suggests a variation of [\ion{N}{2}]/H$\alpha$ due to an abundance effect. This
behavior is consistent with our interpretation of Figure~3, and suggests 
that this AGN probably had a starburst in its outer region (like in HCG16-1, for example)  
in the recent past. 

The values observed in the starburst HCG16-3 are consistent with excitation
produced by massive stars. The outer regions however show values that could be interpreted
as the products of shocks. The same behavior is observed in HCG16-4, although at a much
lower level. This is consistent with the idea that HCG16-4 is much more active 
than HCG16-3. In this galaxy the burst population in the outer regions, though more evolved than
in the nucleus, are however younger than in the outer regions of HCG16-3. 

Again, the analysis of HCG16-5 is the most complex. The values for the primary nucleus
are at the lower limit for AGN and starburst and are consistent with shocks. The secondary
nucleus has values consistent with shocks and AGN. All the outer regions show values unusually
high, suggesting the presence of shocks or domination by an AGN. This observation
supports our previous interpretation 
that HCG16-5 is a mixture of two AGNs with starbursts in their nucleus.
  
\subsection{Variation of the stellar populations with the radius}

In this section we complete our analysis for our 5 galaxies
by studying the characteristics of their stellar populations, as deduced from
the absorption features. For this study, 
we measured the absorption features in three apertures. The results
are presented in Table 3. The three apertures are the same as those 
used for the activity classification. The corresponding widths in kpc
are given in column 3. The absorption features were measured by drawing 
a pseudo continuum by eye using a region $\sim 100$ \AA\ wide on each side of the line.
Columns 4 to 10 give the EW of the most prominent absorption features in the spectra.
Column 11 gives the ratios of the center of the line intensity of the \ion{Ca}{2} H +
H$\epsilon$ lines to the center of the line intensity of the \ion{Ca}{2} K and 
column 12 gives the Mg$_2$ index. The uncertainties were determined the 
same way as for the emission line features. 

In Figure~5, we show the diagram of the EW of H$\delta$ as a function of 
the (\ion{Ca}{2} H + H$\epsilon$)/\ion{Ca}{2} K index (Rose 1985). 
This diagram is useful for identifying post-starburst galaxies (Rose 1985; Leonardi \& Rose 1996;
Poggianti \& Barbaro 1996; Zabludoff et al. 1996; Caldwell et al. 1996; Barbaro \& Poggianti 1997;
Caldwell \& Rose 1997).  
Galaxies with intermediate age populations have high EW of H$\delta$ and high values
of the (\ion{Ca}{2} H + H$\epsilon$)/\ion{Ca}{2} K ratios. 
From this diagram, it can be seen that the five galaxies 
in HCG 16 show the presence of intermediate age stellar populations. 

In Figure~5, we compare the five galaxies in HCG 16 with
the sample of HCG galaxies previously studied by Coziol et al. (1998b). It can 
be seen that the five galaxies in HCG 16 have characteristics which indicate
younger post-starburst phases than in most of the galaxies in Coziol et al. (1998b). 
This observation is consistent with our scenario for the formation of the
groups, which suggests that HCG 16 is an example of a young group. 

In Figure~5, it is interesting to compare the position of the two starburst
galaxies HCG16-3 and HCG16-4.
The position of HCG16-3 suggests that it contains more
intermediate age stars than HCG16-4. But at the same time we deduce 
from Figure~3 that HCG16-4 has a younger burst than  
HCG16-3. How can we understand this apparent contradiction? 
One possibility is to assume that the EW(H$\delta$) in HCG16-4 is contaminated
by emission, explaining the low EW observed for this galaxy.
For the (\ion{Ca}{2} H + H$\epsilon$)/\ion{Ca}{2} K
indices we note also that these values are comparable 
with those produced by very massive stars (Rose 1985). 
Another alternative, however, would be to suppose that the stellar populations 
are from another generation
suggesting multiple bursts of star formation in HCG16-4 (Coziol 1996; Moore, Lake, \& Katz 1998;
Smith et al. 1998; Taniguchi \& Shioya 1998).

In Figure~5, the position of HCG16-2 is consistent with no star formation in its
nucleus. It could have been higher in the outer regions in the recent past, which is consistent with
our interpretation of Figures~3 and 5 for this galaxy.
We also note the very interesting position of HCG16-5, which shows
a strong post-starburst phase in the two nuclei and in the outer regions. 
This observation supports our previous interpretation 
of these two LINERs as a mixture of AGNs with starbursts in their nuclei. 

Finally, we examine the stellar metallicities of our galaxies, as deduced from the Mg2 index
(Burstein et al. 1984; Brodie \& Huchra 1990; Worthey, Faber, \& Gonz\'alez 1992; Bender, Burstein,
\& Faber 1993).
In Figure~6, the stellar metallicity is shown as
a function of the ratio EW(\ion{Ca}{2} H + H$\epsilon$)/EW(\ion{Ca}{2} K), which
increases as the stellar population get younger (Rose 1985; Dressler \& Schectman 1987). 
For our study, we assume that a high value of the Mg2 index 
indicates a high stellar metallicity. In Figure~6, the range of Mg2 generally observed
in late type spirals is indicated by two dotted lines. The upper limit for the
early--type galaxies is marked by a dashed line.  

Figure~6 suggests that, 
the stellar populations are generally more metal rich in the nuclei than in the 
circumnuclear regions. The two AGNs, HCG16-1 and HCG16-2, are more metal rich, and, therefore, 
more evolved. HCG16-3 and HCG16-4 have, on the other hand, typical values for starburst galaxies
(Coziol et al. 1998). In terms of stellar population and metallicity HCG16-5 is more similar
to HCG16-3 and HCG16-4, which suggests a similar level of evolution. 

\section{Discussion}

Our observations are consistent with the existence of a close relation between 
AGN and starbursts. In our sample the most obvious case is HCG16-1, which has 
a LINER nucleus and star formation in its outer regions.
A similar situation was probably present in HCG16-2, in a recent past.  
HCG16-5, on the other hand, shows a very complicated case where we cannot
clearly distinguish between star formation an AGN. 
The question then is what is the exact relation between these two phenomena?

One possibility would be to assume that AGN and starburst are, 
in fact, the same phenomenon (Terlevich et al. 1991):
the AGN characteristics are produced by the evolution of a massive starbursts
in the center of the galaxies. HCG16-5 could be a good example of this. 
However, nothing in our observations
of this galaxy allows us to identify the mechanism producing the LINER 
with only star formation. In fact, the similarity of HCG16-5 to HCG16-2 suggests
that what we see is more a mixture of the two phenomena,   
where an AGN coexists in the nucleus with a starburst (Maoz et al. 1998;
Larkin et al. 1998; Gonzalez-Delgado et al. 1997; Serlemitsos, Ptak, \& Yaqoob 1997). 

Perhaps the two phenomena are different, but still related via 
evolution. In one of their recent paper, Gonzalez-Delgado et al. (1997) proposed a 
continuous sequence where a starburst is related to a Seyfert 2,
which, at the end, transforms into a Seyfert 1.
Following our observations, it is interesting to see that in terms of stellar populations,
HCG16-1 and  HCG16-2 are the most evolved galaxies of the group. In Coziol et al. (1998b)
we also noted that this is usually the case for the luminous AGN and low--luminosity
AGN galaxies in the groups. The AGNs in the samples of
Gonzalez-Delgado et al. (1998) and in Hunt et al. (1997) all look like
evolved galaxies. However, as we noted in the introduction, 
we have not found any Seyfert 1 in the 60 compact groups we have investigated (Coziol et al. 1998). 
Following the scenario of Gonzalez-Delgado et al. (1998) this would simply
mean that the groups are not evolved enough. This is difficult
to believe as it would suggest that we observe all these 
galaxies in a very special moment of their existence. 
In Coziol et al. (1998b) the observations suggests
that the end product of the evolution of the starburst--Seyfert 2 connection in the groups 
is a low--luminosity AGN or a galaxy without emission lines. 

Maybe, there are no Seyfert 1 in the groups because the conditions 
for the formation of these luminous AGNs are not satisfied in the groups. On this matter,
it is interesting to find two mergers in HCG 16: HCG16-4 and HCG16-5. But galaxy
HCG16-4 is a strong starburst while HCG16-5 is, at most, a LINER or a Seyfert 2.
Could it be then that the masses of
these two mergers were not sufficient to produce a Seyfert 1? Maybe
the mass of the merging galaxies and/or the details on how 
the merging took place are the important parameters (Moles, Sulentic, \& M\'arquez 1998;
Moore, Lake, \& Katz 1998; Lake, Katz, \& Moore 1998; Taniguchi 1998; Taniguchi \& Shioya 1998). 

An evolutionary scenario for the starburst--AGN connection
is probably not the only possible alternative. It could also be that
the presence of a massive black hole (MBH) in the nucleus of an AGN
influences the evolution of the star formation (Perry 1992; Lake, Katz, \& Moore 1998;
Taniguchi 1998). One can imagine, for instance,
that a MBH is competing with the starburst for the available gas.
Once the interstellar gas has become significantly concentrated 
within the central region of the galaxy, it could accumulates
in an extended accretion disk to fuel the MBH. Assuming 10\%
efficiency, accretion representing only 7 M$_\odot$\ yr$^{-1}$\ will easely 
yield 10$^{13}$\ L$_\odot$, while astration rates of 10--100 M$_\odot$\ yr$^{-1}$\
are necessary to produce $10^{11}-10^{12}$\ L$_\odot$ (Norman \& Scoville 1988). 
Obviously the gas that goes into the nucleus to feed the MBH will not be available
to form stars, hence the star formation phase will have a shorter lifetime. 
Other phenomena also related to AGNs, like jets, ejection of gas, or even
just a very extended ionized region could stimulate or inhibit star formation
in the circumnuclear regions (Day et al. 1997; Falcke 1998; Quillen \& Bower 1998). 
Obviously, the more active the AGN
the greater its influence should be. Therefore, the fact that most of
the AGNs in the compact groups are of the shallower types (Seyfert 2, LINER and
low--luminosity AGN) suggests that these phenomena probably were not so important
in the groups.

Another interesting aspect of our observations concerns the origin of the compact groups.
In Coziol et al. (1998b) and Ribeiro et al. (1997), we suggest that 
the core of the groups are old
collapsing structures embedded in more extended systems where they are replenished in
galaxies (Governato, Tozzi, \& Cavaliere 1996). We have also proposed an evolutionary scenario for the formation
of the galaxies in the group. Following this scenario, HCG 16
would be an example of a group at an early stage of its evolution.  
Our present observations support this scenario and give us further
insights on how the groups could have formed. 

The original core of HCG 16 is formed of the galaxies HCG16-1, HCG16-2, HCG16-4 and 
HCG16-5 (Ribeiro et al. 1997).
Our observations now suggest that HCG16-1, HCG16-2 form the evolved core 
of HCG 16, while HCG16-4 and HCG16-5 are more recent additions. 
The fact that we see traces of mergers in these two last galaxies suggests
that HCG16-4 and HCG16-5 originally were not two
massive galaxies but 4 smaller mass, metal poor galaxies.  
The remnant star formation activity in HCG16-1, HCG16-2 could also indicate that they too 
were formed by mergers, but a much longer time ago. This scenario may 
resolve the paradox of why galaxies in the cores of the HCGs have not already 
merged to form one big galaxy (Zepf \& Whitmore 1991; Zepf 1993). If HCG 16 is typical of
what happened in the other groups, then originally the number of galaxies was higher
and their mass lower and hence the dynamic of the groups was much different.
HCG16-3 looks, on this matter, like a more recent addition, and suggests
that the process of formation of the group is still going on today.

\acknowledgments  

We would like to thank  Roy Gal, and Steve Zepf for very useful suggestions.

\clearpage 

\figcaption[f1.ps]{Extension of the ionized gas centered at H$\alpha$
and light profiles along the slit. The direction of the east
is indicated. At the left, the extension in kpc of the region of
the spectra with 90\% of the light is indicated. This same region is marked 
in the light profile by a dashed line at the 10\% level of intensity.
The FWHM and total extension of the galaxies are given in Table~1.
The profiles of HCG16-4 and HCG16-5 show secondary peaks corresponding to secondary nuclei.}

\figcaption[f2.ps]{Standard diagnostic diagram of line ratios as measured
in three different apertures. The value for the nuclei are identified by filled symbols.
The horizontal dot line separate Seyfert 2 (and HII galaxies) from LINER (SBNGs).
The continuous curve is the empirical separation between AGN and starburst as given
by Veilleux \& Osterbrock (1987).}

\figcaption[f3.ps]{[\ion{N}{2}]/H$\alpha$ ratios as a function of the
EW of H$\alpha$, as measured using 7 equal apertures of $\sim 3$ arc seconds. 
The nuclei are identified by crosses and the circumnuclear region 
by a small filled dot. The numbers correspond to the different apertures as given in
Table~2. We do not display any error bar because they are quite small, comparable to the
size of the symbols.}

\figcaption[f4.ps]{[\ion{N}{2}]/H$\alpha$ ratios as a function of the
[\ion{S}{2}]/H$\alpha$, as measured using 7 equal apertures of $\sim 3$ arc seconds. 
The numbers correspond to the different apertures as given in
Table~2.}

\figcaption[f5.ps]{The EW of H$\delta$ line in function of the 
(Ca II + H$\epsilon$)/Ca II K index. 
The horizontal dashed line separate
post-starbursts from normal late--types spirals (see Coziol et al. 1998b).} 

\figcaption[f6.ps]{The Mg2 index in function of the ratio of the
EW of the Ca II + H$\epsilon$ and Ca II K lines. Range values of Mg2 for 
late--type spirals is indicated by the two horizontal dotted lines. The horizontal
dashed lines is the higher limit for normal early--type galaxies.}

\clearpage

\begin{deluxetable}{lccclcccc}
\footnotesize
\tablecaption{Characteristics of the galaxies in the group\label{tbl-1}}
\tablewidth{0pt}
\tablehead{
\colhead{HCG  \#} &\colhead{cz}            & \colhead{M$_{\rm B}$} & \colhead{T} &
\colhead{Activity}&\multicolumn{3}{c}{Extension of light in the spectra}         &\colhead{1 arcsec }     \\
\colhead{}        &\colhead{}              & \colhead{}            & \colhead{}  &
\colhead{Type}    &\colhead{Total galaxy}& \colhead{Ionized regions}& \colhead{Nucleus}& \colhead{}\\
\colhead{}        &\colhead{(km s$^{-1}$)} & \colhead{}            & \colhead{}  &
\colhead{}        &\colhead{(arcsec)}    & \colhead{(arcsec)}       & \colhead{(arcsec)}& \colhead{(parsec)}\\ 
}
\startdata
16  01  &4073   &-20.79&      2 &  LNR$+$ SBNG &  36 &32 & 1.8 &263\nl 
16  02  &3864   &-20.21&      2 &  Seyfert 2$+$LNR   &  31 &21 & 3.7 &250\nl 
16  03  &4001   &-20.29&\nodata &  SBNG        &  34 &28 & 3.7 &259\nl 
16  04  &3859   &-19.95&    10  &  SBNG        &  45 &40 & 6.0 &249\nl 
16  05  &3934   &-19.94&\nodata &  LNR$+$Seyfert 2   &  40 &15 & 8.3 &254\nl 
\enddata 
\end{deluxetable}

\clearpage
\begin{deluxetable}{llccccc}
\footnotesize
\tablecaption{Variation of the emission characteristics as a function of the radius\label{tbl-2}}
\tablewidth{0pt} 
\tablehead{
\colhead{HCG  \#} &\colhead{Ap. \#}& radius & \colhead{FWHM}        &\colhead{[NII]/H$\alpha$} &
\colhead{EW} & \colhead{[SII]/H$\alpha$} \\
\colhead{}        &\colhead{}&\colhead{(kpc)}      &\colhead{(km s$^{-1}$)}& \colhead{}                            & 
\colhead{(\AA)}        & \colhead{} 
}
\startdata
16  01 &1  &+2.37&  \nodata    & 0.6 $\pm$0.1 &   14           &  0.31$\pm$0.04 \nl
       &2  &+1.58&  \nodata    & 0.60$\pm$0.01&   13           &  0.31$\pm$0.01 \nl
       &3ci&+0.79&  118$\pm$  1& 1.4 $\pm$0.2 &    4$\pm$1     &  0.8 $\pm$0.1  \nl
       &4n & 0   &  115$\pm$  6& 1.4 $\pm$0.2 &    4$\pm$1     &  0.7 $\pm$0.1  \nl
       &5ci&-0.79&  112$\pm$ 16& 1.4 $\pm$0.2 &    4$\pm$1     &  0.7 $\pm$0.1  \nl
       &6  &-1.58&  \nodata    & 0.52         &   20$\pm$1     &  0.27$\pm$0.02 \nl
       &7  &-2.37&  126$\pm$ 14& 1.0 $\pm$0.2 &    3$\pm$1     &  0.9 $\pm$0.2  \nl
16  02 &1  &+2.25&  \nodata    & 2.7 $\pm$0.2 &    1.2$\pm$0.1 &  1.8 $\pm$0.8  \nl
       &2  &+1.50&  \nodata    & 2.1 $\pm$0.1 &    2.1$\pm$0.1 &  1.0 $\pm$0.2  \nl
       &3ci&+0.75&  539$\pm$  3& 2.1 $\pm$0.1 &    3.9$\pm$0.3 &  1.0 $\pm$0.1  \nl
       &4n & 0   &  522$\pm$ 12& 1.9 $\pm$0.2 &    4.3$\pm$0.4 &  1.7 $\pm$0.1  \nl
       &5ci&-0.75&  547$\pm$ 14& 1.9 $\pm$0.1 &    4.5$\pm$0.2 &  1.8 $\pm$0.1  \nl
       &6  &-1.50&   89$\pm$  7& 3.1 $\pm$0.2 &   0.89$\pm$0.04&  2.4 $\pm$0.2  \nl
16  03 &2  &+1.55&  \nodata    & 0.60$\pm$0.03&    11  $\pm$2  &  0.61$\pm$0.03 \nl
       &3ci&+0.78&  \nodata    & 0.45         &    34.5$\pm$0.1&  0.4 $\pm$0.1  \nl
       &4n & 0   &  \nodata    & 0.45         &    34.5$\pm$0.2&  0.36          \nl
       &5ci&-0.78&  \nodata    & 0.45$\pm$0.01&    34.5$\pm$0.2&  0.36$\pm$0.01 \nl
       &6  &-1.55&  \nodata    & 0.68$\pm$0.02&     6.9$\pm$0.4&  0.60$\pm$0.04 \nl
16  04 &1  &+2.24&  \nodata    & 0.42$\pm$0.01&   55$\pm$2     &  0.36$\pm$0.01 \nl
       &2  &+1.49&  \nodata    & 0.39         &   89$\pm$2     &  0.30          \nl
       &3ci&+0.75&  \nodata    & 0.39         &   95$\pm$2     &  0.28$\pm$0.01 \nl
       &4n & 0   &  \nodata    & 0.39         &  126$\pm$3     &  0.23          \nl
       &5ci&-0.75&  \nodata    & 0.39         &  122$\pm$1     &  0.24          \nl
       &6  &-1.49&  \nodata    & 0.46$\pm$0.01&   43$\pm$1     &  0.46          \nl
16  05 &1  &+3.05&  \nodata    & 2.49$\pm$0.07&   1            &  1.41$\pm$0.01 \nl
       &2  &+2.29&  \nodata    & 2.0 $\pm$0.2 &   1.4$\pm$0.3  &  1.5 $\pm$0.3  \nl
       &3  &+1.53&  \nodata    & 1.38$\pm$0.02&   2.9$\pm$0.2  &  0.8 $\pm$0.1  \nl
       &4ci&+0.76&  \nodata    & 0.67$\pm$0.01&  16.6$\pm$0.1  &  0.51$\pm$0.01 \nl
       &5n & 0   &  \nodata    & 0.61         &  46  $\pm$2    &  0.38$\pm$0.01 \nl
       &6ci&-0.76&  \nodata    & 0.61         &  46  $\pm$1    &  0.38$\pm$0.01 \nl
       &7  &-1.53&  \nodata    & 1.07$\pm$0.07&   3.4$\pm$0.4  &  1.0 $\pm$0.1  \nl
\tablecomments{Apertures spanning the nucleus and the circumnuclear regions
are indicated by n and ci respectively}
\tablecomments{Radius are positive to the east and negative to the west} 
\enddata 
\end{deluxetable}

\clearpage
\begin{deluxetable}{llccccccccc}
\footnotesize
\tablecaption{Variation of the absorption features with the aperture \label{tbl-3}}
\tablewidth{0pt}
\tablehead{
\colhead{HCG  \#}& \colhead{Ap. \#}& \colhead{Width}&\colhead{CaII K} & \colhead{CaII H} & \colhead{H$\delta$} & \colhead{G--band} &
 \colhead{H$\beta$}  & \colhead{$\frac{{\rm I(CaII H)}}{{\rm I(CaII K)}}$}      & \colhead{Mg$_2$}  \\
\colhead{}       & \colhead{}      & \colhead{(kpc)}   & \colhead{(\AA)} & \colhead{(\AA)}  & \colhead{(\AA)}     & \colhead{(\AA)}   & 
 \colhead{(\AA)}     & \colhead{ }                                              & \colhead{ }  
}
\startdata
16  01&1  &1.58&   5.0$\pm$0.7  &  6.9$\pm$0.3  &  4.4$\pm$1.0 &4.2$\pm$0.4& 5.0$\pm$0.2& 0.84$\pm$0.04& 0.266$\pm$0.004\nl
      &2n &0.53&   7.0$\pm$0.4  &  7.27$\pm$0.02&  3.8$\pm$0.2 &5.1$\pm$0.6& 5.1$\pm$0.4& 0.88$\pm$0.04& 0.337$\pm$0.002\nl
      &3  &2.10&   5.5$\pm$0.4  &  5.1$\pm$0.8  &  3.3$\pm$0.4 &5.2$\pm$0.4& 5.8$\pm$0.1& 1.03$\pm$0.02& 0.29$\pm$0.01  \nl
16  02&1  &2.50&  16  $\pm$2    & 10.1$\pm$1.7  &  3.0$\pm$0.2 &6.3$\pm$3.4& 3.6$\pm$2.6& 0.84$\pm$0.08& 0.3$\pm$0.2    \nl
      &2n &0.50&  12.3$\pm$0.2  &  8.5$\pm$0.3  &  1.4$\pm$0.1 &8.4$\pm$0.2& 1.6$\pm$0.1& 1.11$\pm$0.05& 0.39$\pm$0.01  \nl
      &3  &2.75&  14.9$\pm$0.5  &  7.9$\pm$0.5  &  2.4$\pm$0.5 &3.1$\pm$0.9& 3$\pm$2    & 0.96$\pm$0.07& 0.2$\pm$0.2    \nl
16  03&1  &1.81&   6.8$\pm$0.1  & 10.8$\pm$0.8  & 12.0$\pm$0.5 &\nodata    &11$\pm$1    & 0.66$\pm$0.06& 0.19$\pm$0.04  \nl
      &2n &0.52&   4.2$\pm$0.7  &  8$\pm$1      &  7$\pm$1     &\nodata    & 7$\pm$1    & 0.88$\pm$0.07& 0.18$\pm$0.03  \nl
      &3  &1.55&   3.3$\pm$0.1  &  8$\pm$1      &  7$\pm$1     &\nodata    & 7.7$\pm$0.9& 0.8$\pm$0.1  & 0.15$\pm$0.01  \nl
16  04&1  &1.00&   1.6$\pm$0.7  &  3.2$\pm$0.6  &  2.6$\pm$0.4 &\nodata    & \nodata    & 0.95$\pm$0.06& 0.11$\pm$0.02  \nl
      &2n &0.50&   1.4$\pm$0.3  &  3.2$\pm$1.1  &  2.7$\pm$0.4 &\nodata    & 2.5$\pm$0.6& 1.02$\pm$0.03& 0.125$\pm$0.002\nl
      &3  &2.24&   1.4$\pm$0.1  &  3.9$\pm$0.4  &  3.4$\pm$0.7 &\nodata    & 3.0$\pm$0.1& 0.98$\pm$0.03& 0.114$\pm$0.004\nl
16  05&1n2&2.79&   5.8$\pm$0.8  & 12$\pm$1      &  9.3$\pm$1.2 &3.0$\pm$0.2& 9.3$\pm$0.7& 0.72$\pm$0.07& 0.13$\pm$0.05  \nl
      &2n1&1.02&   3.75$\pm$0.01&  9$\pm$1      &  8.4$\pm$1.2 &3.4$\pm$1.7& 7.7$\pm$0.5& 0.82$\pm$0.02& 0.166$\pm$0.003\nl
      &3  &1.78&   4.1$\pm$0.8  &  9.3$\pm$0.8  &  7.7$\pm$0.2 &3.1$\pm$0.2& 7.9$\pm$0.3& 0.79$\pm$0.03& 0.16$\pm$0.01  \nl
\enddata 
\end{deluxetable}

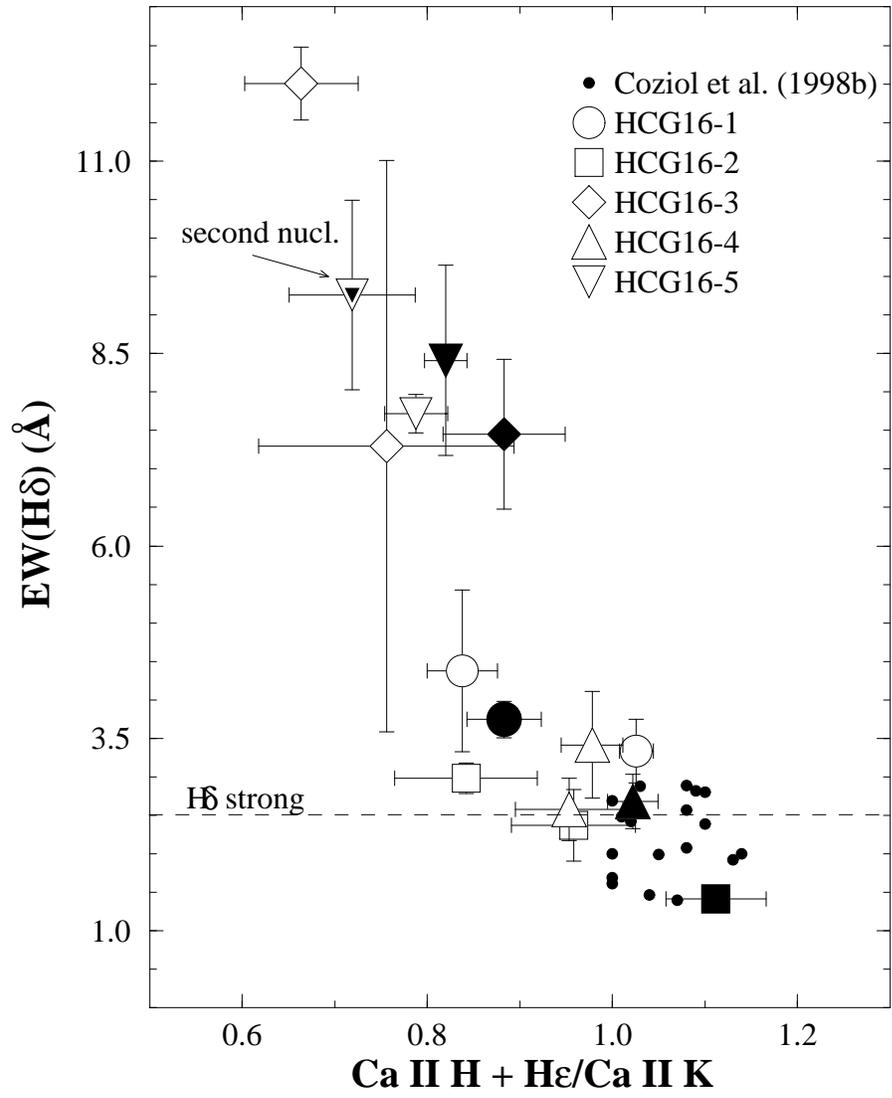
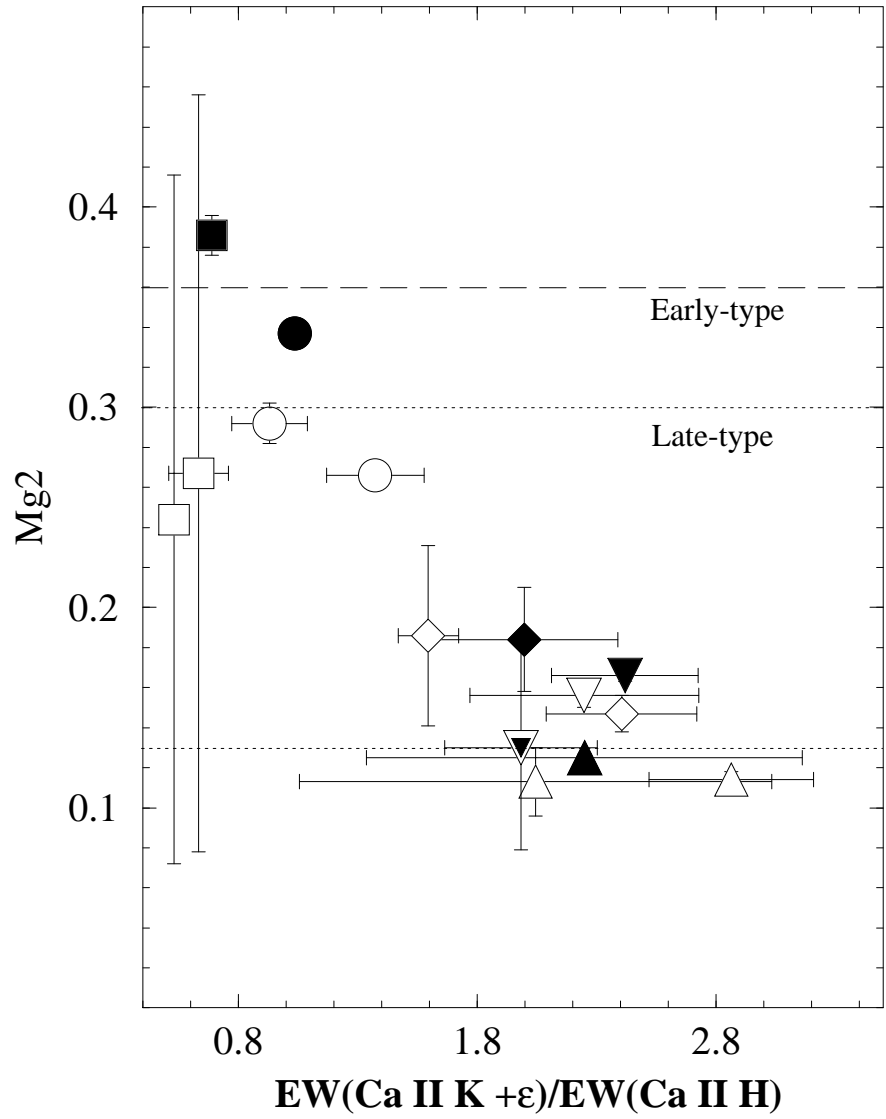
\end{document}